\documentclass[12pt,a4paper]{article}
\usepackage{amsmath}
\usepackage[T1]{fontenc}
\usepackage[utf8]{inputenc}
\usepackage{authblk}

\usepackage{hyperref}
\usepackage{graphicx}
\usepackage{fancybox}

\def\G{\Gamma}
\def\D{\Delta}

\def\a{\alpha}
\def\b{\beta}

\def\d{\delta}
\def\e{\epsilon}
\def\m{\mu}
\def\n{\nu}
\def\s{\sigma}
\def\r{\rho}

\def\o{\omega}

\def\p{\partial}

\title{\vspace{-2cm}
       \hfill
       \mbox{\normalsize{NORDITA-2012-81}}
       \vskip 20pt
Chiral primary one-point functions in the D3-D7 defect conformal field theory}

 \author[1]{Charlotte Kristjansen}
\author[1,2]{Gordon W. Semenoff}
\author[3]{Donovan Young}
 
\affil[1]{Niels Bohr Institute, Copenhagen University,
Blegdamsvej 17,   2100 Copenhagen \O, Denmark}
\affil[2]{Department of Physics and Astronomy, University of British Columbia,  Vancouver, BC Canada V6T 1Z1}
\affil[3]{Nordita, KTH Royal Institute of Technology and Stockholm University,
Roslagstullsbacken 23, SE-106 91 Stockholm, Sweden}

\date{} 
\begin{document}

\maketitle

\begin{abstract}
We compute the one-point functions of chiral primary operators in the non-supersymmetric 
defect conformal field theory
that is dual to the IIB string theory on $AdS_5\times S^5$ background with a probe D7 brane with internal
gauge field flux, both in perturbative Yang-Mills theory and in the string theory dual. The former is expected to
be accurate at weak coupling whereas the latter should be 
accurate in the planar strong coupling limit of the gauge theory. 
We consider the distinct cases where the D7 brane has geometry $AdS_4\times S^4$ with an instanton bundle
of the worldvolume gauge fields on $S^4$ and $AdS_4\times S^2\times S^2$ with Dirac monopole bundles
on each $S^2$.  The gauge theory computation and the string theory computation can be compared directly 
in the planar limit and then a subsequent limit where the 
worldvolume flux is large.
We find that  there is exact agreement between the two in the leading order.  
\end{abstract}

\section{Introduction}

Probe branes and defect conformal field theories have been widely
studied in the context of AdS/CFT holography
\cite{Karch:2001cw}-\cite{Grignani:2012qz}.  The classic example is
the D3-D5 intersection \cite{DeWolfe:2001pq,Erdmenger:2002ex}
where the relative orientation of
the D3 and D5 branes is shown in table 1. In the appropriate
limit, this system becomes $AdS_5\times S^5$ bisected by the probe D5 brane with
world volume geometry $AdS_4\times S^2$.  The original idea 
\cite{Karch:2001cw,Karch:2000gx,Karch:2000ct} was that, 
in a sense, the AdS/CFT correspondence acts twice. 
The near horizon geometry of the D3-D5 system contains two sets of
excitations, those of closed IIB superstrings occupying $AdS_5$, which
are dual to ${\mathcal N}=4$ supersymmetric Yang-Mills theory on the $R^4$
boundary of $AdS_5$, 
and open strings connecting the D3 and D5 branes, which are dual to
field theory excitations on the $R^3$ boundary of $AdS_4$, the latter forming a co-dimension one defect in
$R^4$.  
The resulting field theory dual is then ${\cal N}=4$
supersymmetric Yang-Mills theory occupying 3+1 dimensions and interacting with a
field theory living on a 2+1-dimensional defect. The D3-D5 system preserves half of the
supersymmetries of the bulk ${\mathcal N}=4$ theory and 
the degrees of freedom which live on the defect
are a bi-fundamental hypermultiplet which transforms in the fundamental
representation of the bulk gauge group and the fundamental representation of
the defect gauge group.  The defect theory  
reduces the conformal symmetry of ${\cal N}=4$ Yang-Mills theory to
the $SO(2,3)$ conformal symmetry of the defect.  It also reduces the 
$SO(6)$ R-symmetry to $SO(3)\times SO(3)$ and the superalgebra from $PSU(2,2|4)$ to
$OSP(4|4)$. 

\begin{align}\label{dbranes}\nonumber
\boxed{\begin{array}{rcccccccccccl}
  & & x^0 & x^1 & x^2 & x^3 & x^4 & x^5 & x^6 & x^7 & x^8 & x^9 &\\
& D3 & \times & \times & \times & \times & & &  & & & & \\
& D5 & \times & \times & \times &  & \times  & \times & \times & &  & &  \\
& D7 & \times & \times & \times &  & \times  & \times & \times & \times & \times & &   \\
\end{array}}
\nonumber
\\
{\rm\bf Table~1:~D3,~ D5~ and~ D7~orientation}~~~~~~~~~~~~~~~~~~~~~~~~
\nonumber
\end{align}

\begin{figure}\begin{center}
 \includegraphics[bb=19 26 417 366, width=3in]{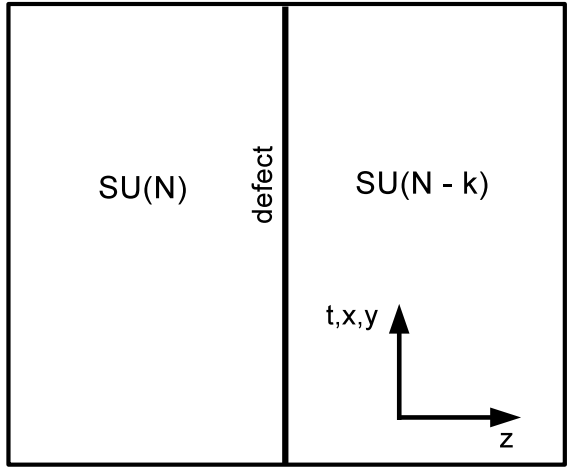}\\
\end{center}
\caption{{\bf Defect conformal field theory}: The defect is 
depicted by the vertical line through the center of the diagram.   
The defect occupies 2+1-dimensions and divides the 3+1-dimensional
space-time into two different regions which are occupied by ${\cal
  N}=4$ supersymmetric Yang-Mills theories with gauge group $SU(N)$ on
the left and $SU(N-k)$ on the right. The integer $k$ is the number
of D3 branes which end on the D5 or D7 brane.  In the latter cases, $k$
should be replaced by $k_1k_2$ or $d_G$.}
\label{dbranes2}\end{figure}

An interesting variant of the D3-D5 system is where the worldvolume
gauge fields of the probe D5 brane have a monopole bundle with a
quantized U(1) magnetic flux.  The geometry of the brane is still
$AdS_4\times S^2$ and the flux is on the $S^2$.  As depicted in
Figure \ref{dbranes2}, in the field theory dual, the defect separates
3+1-dimensional space-time into regions where the gauge group of
${\mathcal N}=4$ Yang-Mills theory has different ranks, $N$ on one
side of the defect and $N-k$ on the other side of the defect, with $k$
the number of units of Dirac monopole flux.  In the string theory this
corresponds to the situation where $k$ of the N D3 branes end on the
worldvolume of a D5 brane.  The low energy action of the D5 brane
contains a Wess-Zumino term proportional to $\int_{D5} C^{(4)}\wedge
F$, with $C^{(4)}$ the Ramond-Ramond 4-form carrying the D3 brane
charge, and $F$ the field strength of the worldvolume gauge field.
The D3 branes intersect the D5 brane on a subvolume which can be
linked by a 2-sphere.  The first Chern class of the gauge field, the
integral of $F$ over the 2-sphere must be equal to the number of
$D3$ branes which terminate there.  The holographic system with this
flux has some interesting new features which have been exploited in
holographic constructions of 2+1-dimensional physical systems
\cite{Myers:2008me,Grignani:2012jh}.  In the limits which are taken in
holographic theories, the magnetic flux has to be large,
$k\sim\sqrt{\lambda}$, where $\lambda$ is the 't Hooft coupling of
${\mathcal N}=4$ Yang-Mills theory, in order to have an effect on the
geometry of the D5 brane and to be visible in the physics of the
holographic dual.

In an interesting recent paper \cite{Nagasaki:2012re}, Nagasaki and
Yamaguchi showed that the one-point functions of chiral primary
operators could be computed for the D3-D5 defect field theory in the
limit of large monopole number $k$.  The defect preserves an
$SO(3)\times SO(3)$ subgroup of the $SO(6)$ R-symmetry and, in the
defect conformal field theory, an $SO(3)\times SO(3)$ invariant
primary operator ${\mathcal O}_\Delta(x)$ can have a one-point
function
\begin{align}
\left<{\mathcal O}_\Delta(x)\right> = \frac{C_\Delta} {|z|^\Delta},
\end{align}
where $\Delta$ is the conformal dimension of the operator in the bulk
and $|z|$ is the distance between $x$ and the defect.  The result for
the one-point function of the   chiral primary operator 
with conformal dimension $\Delta$  in the limit of  large $\frac{k}{\sqrt{\lambda}}$ 
\cite{Nagasaki:2012re} is
\begin{align}\label{d5onepointfunction}
\left< {\mathcal O}_\Delta(x)\right> = 
\frac{ k}{\sqrt{\Delta}} \left(\frac{2\pi^2k^2}{\lambda}\right)^{\Delta/2} Y_\Delta(0) ~\frac{1}{|z|^\Delta},
\end{align}
where $\Delta$ must be an even integer and   $Y_\Delta (\psi)$ is the unique $SO(3)\times SO(3)$ symmetric spherical harmonic
at level $\Delta$.  (See Section 2, equation (\ref{sph})). In equation (\ref{d5onepointfunction}), it is evaluated at $\psi=0$, where
the $S^2$ has maximum volume. 

What is remarkable is that this result is obtained on both the gauge
theory and the string theory side.  In both cases, it is of leading order
in the parameter $\frac{\sqrt{\lambda}}{k}$ which is assumed to be
small on both sides. It is one of the few
instances where the gauge and string theory computations can be
compared directly.  Moreover, it is a test of the AdS/CFT
correspondence in the presence of probe branes complementary to 
previous tests which have examined the open D3-D5 string sector in the
plane wave BMN limit \cite{Berenstein:2002jq} \cite{Berenstein:2002zw}-\cite{DeWolfe:2004zt}.
The accuracy of the BMN limit in the string theory depends on 
a large angular momentum on $S^5$ whereas, in the present case, it is $k$, the number
of D3 branes which end on the D5 brane that is large.

In this Paper, we will extend and elaborate on this interesting
result.  In particular, we will extend it to the non-supersymmetric
D3-D7 system.  We will consider two different cases, the first with 
$SO(3)\times SO(3)$ symmetry
where the probe brane geometry is $AdS_4\times S^2\times
S^2$ and where the worldvolume U(1) gauge field of the D7 brane has $k_1$
and $k_2$ units of magnetic flux on the two 2-spheres, and the second 
with $SO(5)$ symmetry, where
the probe brane geometry has is $AdS_4\times S^4$ and the worldvolume gauge fields
form a certain instanton bundle on the $S^4$ with instanton number $d_G$.
The situation on the gauge theory side is again as illustrated in 
Figure \ref{dbranes2}, only in the first case the parameter $k$  gets replaced by 
$k_1 k_2$ and in the second case it gets replaced by $d_G$.
Remarkably,
even in these non-supersymmetric systems, the one-point functions of
chiral primary operators on the string and gauge theory sides agree 
in
the leading order of a large $\frac{k_1}{\sqrt{\lambda}},\frac{k_2}{\sqrt{\lambda}}$ limit of the $SO(3)\times
SO(3)$ symmetric system,
\begin{align}\label{onepointfunction}
\left< {\mathcal O}_\Delta(x)\right> = \frac{ k_1k_2}{\sqrt{\Delta}}
  \left(\frac{2\pi^2(k_1^2+k_2^2)}{\lambda}\right)^{\Delta/2} Y_\Delta(\arctan(k_2/k_1)) ~\frac{1}{|z|^\Delta},
~~~~(z<0),
\end{align}
or the limit of large $\frac{n}{\sqrt{\lambda}}$, with instanton number $d_G=\frac{1}{6}(n+1)(n+2)(n+3)$, in the $SO(5)$
symmetric system,  
\begin{align}\label{gres0}
\left<{\mathcal O}_\Delta(x)\right> = \frac{n^3}{6\sqrt{\Delta}}\left(\frac{\pi^2 n^2}{\lambda}\right)^{\frac{\Delta}{2}}
{\mathcal Y}_{\Delta}(0)\frac{1}{|z|^\Delta}, ~~~~(z<0),
\end{align}
where ${\mathcal Y}_{\Delta}(\theta)$ is the unique $O(5)$ symmetric spherical 
harmonic at level $\Delta$. (See Section 4.1.1).
Here it is evaluated at $\theta=0$ where the size of the $S^4$ is maximal.
The second remarkable fact is that a discontinuity in what is meant by
chiral primary operator that is expected to occur as one crosses the defect
will also turn out to be 
visible in the supergravity computation.
On the gauge theory side, it is easily seen. 
The Yang-Mills theory computations which result in
(\ref{onepointfunction}) and (\ref{gres0}) are semi-classical, obtained by simply
substituting a classical configuration of the scalar fields of the
${\mathcal N}=4$ Yang-Mills theory.  These fields are non-zero only on
the side of the defect with $SU(N)$ gauge group, they vanish on the side
with $SU(N-k_1k_2)$ or $SU(N-d_G)$ gauge group.  
The Yang-Mills theory therefore
tells us that the leading contribution to the one-point function
(\ref{onepointfunction})  and (\ref{gres0}) is nonzero only on the side of the defect
where the gauge group is larger (thus the qualification $z<0$ in that
formula).  It is interesting to ask whether this is reflected in the
strong coupling, string theory computation.  We shall find that indeed
it is.  There is a discontinuity in the leading order of the one-point
function as the point crosses the D7 brane, with the value being much
smaller on the side of the brane with smaller D3 brane flux.

The D3-D7 system is of interest as a holographic dual of a defect
field theory where the degrees of freedom living on the
2+1-dimensional defect are fermions
\cite{Davis:2008nv}-\cite{Grignani:2012qz}.  The flat space
orientation of the D3 and D7 branes is displayed in Table 1.  This is
an $\#_{ND}=6$ configuration and it is not supersymmetric
\cite{Polchinski:1998rq}.  Nevertheless, in flat space, there is no
tachyon in the spectrum of the D3-D7 open strings and the only zero
modes are in the Ramond sector.  The light degrees of freedom are thus
chiral fermions inhabiting the 2+1-dimensional overlap of the D3 and
D7 worldvolumes.

However, the most straightforward constructions of the D3-D7 system,
where the D7 brane worldvolume is either $AdS_4\times S^4$ or
$AdS_5\times S^2\times S^2$ without fluxes are
unstable~\cite{Davis:2008nv,Rey:2008zz,Myers:2008me,Bergman:2010gm}.
The systems are non-supersymmetric and the D3 branes and D7 brane
repel each other.  The symptoms of the resulting instability are
tachyonic modes which violate the Breitenholder-Freedman bound for
fluctuations of the embeddings.  They are stabilized by adding the
topological fluxes, the instanton bundle on the $S^4$ when the
instanton number is large enough \cite{Myers:2008me} or Dirac monopole
bundles on either one or both of the $S^2$'s in $S^2\times S^2$, when
the flux is large enough \cite{Bergman:2010gm}.  This lower bound on
the flux is, in the first case, where the instanton number (which we
shall call $d_G$) exceeds a number of the order of $\lambda^{3/2}$
and, in the second case, where the monopole numbers $k_1$ and/or $k_2$ are
of order $\sqrt{\lambda}$.  In this Paper, we shall be interested in
the limits where $d_G\gg\lambda^{3/2}$ or $k_1$ and/or $k_2\gg\sqrt{\lambda}$
as it is in this region where the gauge theory and string theory
computations can be compared.

In Section 2, we present the Yang-Mills theory computation of the
one-point function of chiral primary operators in the case where
$SO(3)\times SO(3)$ R-symmetry is preserved.  In Section 3 we present
the same computation in supergravity.  In Section 4 we examine the
case where the $SO(5)$ subgroup of R-symmetry is preserved.  Section 5
contains a discussion and conclusions.

 \section{Defect ${\cal N}=4$ Yang-Mills Theory}
\label{sec:g}
\subsection{Classical solution}

We begin with  a classical solution of  
${\mathcal N}=4$ supersymmetric Yang-Mills theory, where we set the vector
and spinor fields to zero, $A_\mu=0$, $\psi=0$ and look for a solution of the 
remaining equations for the scalar fields, 
\begin{align}
&\nabla^2\phi_i-\sum_{j=1}^6\left[\phi_j,\left[\phi_j,\phi_i\right]\right]=0,
\label{fieldequation1}\\
&\sum_{i=1}^6\left[\phi_i,\nabla\phi_i\right]=0.
\label{fieldequation2}\end{align}
Equation (\ref{fieldequation2}) is the condition that the $SU(N)$ color current is zero.
A solution of these equations 
with the appropriate symmetries is
\begin{align}  \label{classicalsolution1}
\phi_i(z)&=-\frac{1}{z} (t_i^{k_1} \otimes 1_{k_2\times k_2})\oplus
0_{(N-k_1k_2)\times (N-k_1 k_2)},
\hspace{0.7cm} \mbox{for}\hspace{0.3cm} i=1,2,3, \\
\label{classicalsolution2}
\phi_i(z)&=-\frac{1}{z} (1_{k_1\times k_1}\otimes t_i^{k_2})\oplus
0_{(N-k_1k_2)\times (N-k_1 k_2)},
\hspace{0.7cm} \mbox{for}\hspace{0.3cm} i=4,5,6.
\end{align}
Here $t_i^{k_1}$ 
for $i=1,2,3$ are generators of the $k_1$-dimensional irreducible
representation of $SU(2)$ and $t_i^{k_2}$ for $i=4,5,6$ are generators of the
$k_2$-dimensional irreducible representation of $SU(2)$. Furthermore,
$1_{k_1\times k_1}$ and $1_{k_2\times k_2}$ are unit matrices of dimension 
$k_1\times k_1$ and $k_2\times k_2$ respectively. This generalizes the solution that was used in reference~\cite{Nagasaki:2012re} that was suitable for 
a single $S^2$ to a solution suitable for the product $S^2\times S^2$.

The matrices $\phi_i$ have the following property
\begin{eqnarray}\label{classicalnorm1}
\phi_1^2+\phi_2^2+\phi_3^2&=&\frac{1}{4z^2} (k_1^2-1) 1_{k_1k_2\times k_1k_2}
\oplus 0_{(N-k_1k_2)\times(N-k_1k_2)}, \\  \label{classicalnorm2}
\phi_4^2+\phi_5^2+\phi_6^2&=&\frac{1}{4z^2} (k_2^2-1) 1_{k_1k_2\times k_1k_2}
\oplus 0_{(N-k_1k_2)\times(N-k_1k_2)},
\end{eqnarray}
and hence
\begin{eqnarray} \label{classicalnorm3}
{\rm Tr}(\phi_1^2+\phi_2^2+\phi_3^2)&=&\frac{1}{4z^2} (k_1^2-1)\,k_1 k_2, \\
{\rm Tr}(\phi_4^2+\phi_5^2+\phi_6^2)&=&\frac{1}{4z^2} (k_2^2-1)\, k_1 k_2.
\label{classicalnorm4}
\end{eqnarray}
We will use these equations for computing the classical limit of the one-point function
of a chiral primary operator, which we shall define in the next subsection.

\subsection{The Chiral Primary Operators with $SO(3)\times SO(3)$ symmetry}

The chiral primary operators of ${\cal N}=4$ Yang-Mills theory are 
\begin{equation}\label{chiralprimary}
{\cal O}_{\Delta I}(x)\equiv
\frac{(8\pi^2)^{\frac{\Delta}{2}}}
{\lambda^{\frac{\Delta}{2}}\sqrt{\Delta} }C_I^{i_1 i_2\ldots i_{\Delta}}
\,{\rm Tr}\left(\phi_{i_1}(x)\phi_{i_2}(x)\ldots \phi_{i_{\Delta}}(x)\right),
\end{equation}
where the $\phi_i$'s can be any of the six real scalar fields.
Here $\Delta$ is an integer which counts the number of fields and it is also 
equal to the bulk scaling dimension of
the operator which  does not depend on the coupling constant. 
The tensors $C_{I}^{i_1i_2\ldots i_{\Delta}}$ are symmetric
and traceless in the indices $i_1\ldots i_\Delta$ and they  can be chosen
to satisfy an orthogonality relation
\begin{equation}\label{tracelesssymmetric}
\sum_{i_1\ldots i_\Delta=1}^6C_{I_1}^{i_1i_2\ldots i_{\Delta}}C_{I_2}^{i_1i_2\ldots i_{\Delta}}=\delta_{I_1I_2}~~.
\end{equation} 
The indices $I_a$ label different such tensors which are equal in 
number to the
dimension of the   totally symmetric and traceless
irreducible representation of $SO(6)$ with $\Delta$ indices, 
$I_a=1,\ldots,(3+\Delta)(2+\Delta)^2(1+\Delta)/12$.
With the normalization in (\ref{chiralprimary}) and (\ref{tracelesssymmetric}), 
the planar limit of two point functions of the theory without
the defect are unit normalized as 
\begin{equation}\label{chiralprimarynormalization}
\langle
{\cal O}_{\Delta_1 I_1}(x){\cal O}_{\Delta_2I_2}(y)
\rangle=\frac{\delta_{I_1I_2}\delta_{\Delta_1\Delta_2}}{|x-y|^{2\Delta_1}}.
\end{equation}
The irreducible representations of $SO(6)$ and the components of states in each representation are in one-to-one correspondence with
spherical harmonics on $S^5$. Here, we shall follow the notation of 
reference \cite{Lee:1998bxa}.  Using the same tensor as in (\ref{chiralprimary}), a spherical harmonic on $S^5$ can be written as
\begin{equation}
Y_{\Delta I}=C_I^{i_1 i_2\ldots i_{\Delta}} \hat x_{i_1}\hat x_{i_2}\ldots \hat x_{i_\Delta},
\label{tensor}
\end{equation}
where the $\hat x_i$'s are the components of a unit vector coordinate for
the embedding of $S^5$ in $R^6$,
\begin{equation}
\hat x_1^2+\hat x_2^2+\ldots +\hat x_{6}^2=1.
\end{equation} 
Equation (\ref{tensor}) plus the fact that the tensors $C^{i_1 i_2\ldots i_{\Delta}} $ are traceless and symmetric
and the identity $\nabla_j\hat x_i=\frac{1}{|x|}(\delta_{ij}-\hat x_i\hat x_j)$  where $\hat x_i = x_i/|x|$, can easily
be used to show that
\begin{align}
-\vec\nabla^2 Y_{\Delta I} = \frac{\Delta(\Delta+4)}{|x|^2}Y_{\Delta I},
\end{align}
and, remembering that $-\vec\nabla^2=- \frac{1}{x^5}\frac{d}{dx}x^5\frac{d}{dx}+\frac{L_{ij}^2}{x^2}$,
where $L_{ij}=-i(x_i\nabla_j-x_j\nabla_i)$, 
we see that 
$Y_{\Delta I}$ are eigenfunctions of the Laplacian, $L_{ij}^2$, on $S^5$ with eigenvalues
$\Delta(\Delta+4)$.  The index $I$ runs over the $(3+\Delta)(2+\Delta)^2(1+\Delta)/12$ 
linearly independent spherical harmonics at level $\Delta$.  From these degenerate states, 
we must identify those which are symmetric under an $SO(3)\times SO(3)$ subgroup of $SO(6)$. 
Following reference \cite{Nagasaki:2012re}, it is convenient to use coordinates on $S^5$
where
\begin{align}
ds^2=d\psi^2+\cos^2\psi\left(d\theta^2+\sin^2\theta d\phi^2\right) + \sin^2\psi\left(d\tilde\theta^2+\sin^2\tilde\theta d\tilde\phi^2\right).
\end{align}
Here, $S^5$ is constructed as two $S^2$'s fibered over an interval   $\psi\in [0,\frac{\pi}{2}]$.  
Then, knowing that the $SO(3)\times SO(3)$ invariant
spherical harmonic will not depend on any of the coordinates of the $S^2$'s, that is, it will depend only on $\psi$,
it is clear that $SO(3)\times SO(3)$ invariant spherical harmonic, which we shall denote by $Y_\Delta(\psi)$ must obey the equation
\begin{align}\label{differentialequationforyl}
\frac{1}{\cos^2\psi\sin^2\psi}\frac{d}{d\psi}\cos^2\psi\sin^2\psi\frac{d}{d\psi}Y_{\Delta }(\psi)
=-\Delta(\Delta+4)Y_{\Delta }(\psi).
\end{align}
We shall see shortly that, with regularity conditions at $\psi=0$
and $\psi=\pi/2$, there is a non-singular solution of this equation  when $\Delta$ is an even integer. 
 This implies that, when $\Delta$ is even, there is only one
$SO(3)\times SO(3)$ invariant state and when $\Delta$ is odd, there are no 
such states.  

With the change of variables
\begin{equation}\label{definitionofz}
z=e^{2i\psi},
\end{equation}
the differential equation (\ref{differentialequationforyl}) becomes
\begin{equation}\label{homogeneousdifferentialequation}
\left[-\left(z\frac{d}{dz}\right)^2+\left(1+\frac{\Delta}{2} \right)^2 \right]\left[ (z-1/z)Y_\Delta(z)\right]=0,
\end{equation}
which is solved by
\begin{align}\label{yasacharacter}
Y_\Delta(z) &= \frac{(-1)^{\Delta/2}}{ 2^{(\Delta-1)/2}\sqrt{(\Delta+1)(\Delta+2)}}\frac{z^{1+\Delta/2}-z^{-1-\Delta/2}}{z-1/z},~~~\Delta~{\rm even}
\\
&= \frac{1}{ 2^{(\Delta-1)/2}\sqrt{(\Delta+1)(\Delta+2)}}\left( z^{\Delta/2}+z^{\Delta/2-1}+\ldots+z^{-\Delta/2}\right),
\end{align}
where we have chosen the solution which is regular at $z=1$.  It is also regular at $z=-1$ only when $\Delta$
is an even integer, in which case the solution is a polynomial, and we have normalized so that the spherical harmonic
has its usual normalization
\begin{align}\label{normalizationofsphericalharmonic}
\int_{S^5}  |Y_\Delta(\psi)|^2 =
\frac{1}{2^{\Delta-1}(\Delta+1)(\Delta+2)}\int_{S^5} 1.
\end{align}
Using a binomial expansion of $z^{1+\Delta/2} = (\cos\psi+i\sin\psi)^{2+\Delta}$, the spherical harmonic can be presented as an even homogeneous polynomial of order $\Delta$ in $\cos\psi$ and $\sin\psi$,\footnote{The first few of the $SO(3)\times SO(3)$ invariant spherical harmonics are
\begin{align}
Y_0&=1, \nonumber \\
Y_2&=\sqrt{\frac{1}{6}} \left[ \sin^2\psi-\cos^2\psi\right], \nonumber \\
Y_4&= \sqrt{\frac{3}{80}} \left[\sin^4\psi - \frac{10}{3}\sin^2\psi\cos^4\psi+\cos^4\psi\right],
\nonumber \\
Y_6&=\sqrt{\frac{1}{112}}\left[\sin^6\psi-7\sin^4\psi\cos^2\psi+7\sin^2\psi\cos^4\psi
-\cos^6\psi\right],
\nonumber \\
Y_8&= \sqrt{\frac{5}{2304}}  \left[ \sin^8\psi - 12\sin^6\psi\cos^2\psi+\frac{126}{5}\sin^4\psi
\cos^4\psi - 12\sin^2\psi\cos^6\psi + \cos^8\psi\right].
\nonumber
\end{align}
}
\begin{align} \label{sph}
Y_\Delta(\psi)  
&= \frac{    (2+\Delta)! }{  2^{(\Delta+1)/2} \sqrt{(\Delta+1)(\Delta+2)}  }
\sum_{p=0}^{\Delta/2} \frac{(-1)^p \sin^{\Delta-2p}\psi
\cos^{2p}\psi }{ (2p+1)!(1+\Delta-2p)! }.
\end{align}

  When the classical solution (\ref{classicalsolution1}) and 
(\ref{classicalsolution2}) of the classical field equations are substituted into a trace over
gauge group indices such as that in equation (\ref{chiralprimary}), the only components
which survive are those which are invariant under the $SO(3)\times SO(3)$ subgroup
of $SO(6)$.  Therefore, only the components of the chiral primary operator having this symmetry
will have non-vanishing one-point functions.  Since the chiral primary operators are in one-to-one correspondence with the $S^5$ spherical harmonics, and, in the above, we have shown that there is a unique $SO(3)\times SO(3)$
invariant spherical harmonic for each even integer value of $\Delta$, we can conclude that there is one unique $SO(3)\times SO(3)$ symmetric chiral
primary operator for each even $\Delta$.  We shall denote this operator by ${\mathcal O}_\Delta(x)$.  It is a specific linear combination of 
the operators ${\mathcal O}_{\Delta I}(x)$ in a general basis
which we defined in (\ref{chiralprimary}).  We assume that ${\mathcal O}_\Delta(x)$ is normalized as in equation (\ref{chiralprimarynormalization}). 

To relate the trace in (\ref{chiralprimary}) to the spherical harmonic $Y_\Delta$,
we simply factor out the normalization of the six-dimensional vector $(\phi_1,\ldots,\phi_6)$, so that it behaves like a unit
vector.  From equations (\ref{classicalnorm3}) and (\ref{classicalnorm4}) we see that this
results in an overall factor of the normalization to the power of the number of fields $\frac{(k_1^2+k_2^2-2)^{\Delta/2}}{(2|z|)^\Delta}\approx \frac{(k_1^2+k_2^2)^{\Delta/2}}{(2|z|)^\Delta}$ where we shall use the limit where $k_1$ and $k_2$
are large.   Then, 
the trace produces the spherical harmonic $Y_\Delta(\psi)$ where, from (\ref{classicalnorm3}) and (\ref{classicalnorm4}),  we
can identify $\cos\psi=\frac{k_1}{\sqrt{k_1^2+k_2^2}}$ and $\sin\psi=\frac{k_2}{\sqrt{k_1^2+k_2^2}}$ as the relative radii
of the two spheres.  Then, the final trace will produce a factor of $k_1k_2$. 
Putting these
factors together results in the expression (\ref{onepointfunction}) which we recopy here
for the reader's convenience\footnote{
For the first few values of $\Delta$, this expression is
\begin{align}
\left< {\mathcal O}_2(x)\right> &= \sqrt{\frac{1}{12}}   k_1k_2
  \left(\frac{2\pi^2}{\lambda} \right)\left[k_2^2-k_1^2\right]~\frac{1}{|z|^2},
\nonumber
\\
\left< {\mathcal O}_4(x)\right> &= \sqrt{\frac{3}{320 } } k_1k_2 
  \left(\frac{2\pi^2}{\lambda}\right)^{2}\left[k_2^4-\frac{10}{3}k_2^2k_1^2+k_1^4\right] ~\frac{1}{|z|^4},
\nonumber\\
\left< {\mathcal O}_6(x)\right> &=\sqrt{\frac{1}{672 }}   k_1k_2 
  \left(\frac{2\pi^2}{\lambda}\right)^{3} \left[k_2^6-7k_2^4k_1^2+7k_2^2k_1^4
-k_1^6\right] ~\frac{1}{|z|^6},
\nonumber \\
\left< {\mathcal O}_8(x)\right> &=  \sqrt{\frac{5}{18432}}k_1k_2 
  \left(\frac{2\pi^2}{\lambda}\right)^{4} \left[ k_2^8 - 12k_2^6k_1^2+\frac{126}{5}k_2^4k_1^4 - 12k_2^2k_1^6 + k_1^8\right] ~\frac{1}{|z|^8}.
\nonumber 
\end{align}
}
$$ 
\left< {\mathcal O}_\Delta(x)\right> = \frac{ k_1k_2}{\sqrt{\Delta}}
  \left(\frac{2\pi^2(k_1^2+k_2^2)}{\lambda}\right)^{\Delta/2} Y_\Delta(\arctan(k_2/k_1)) ~\frac{1}{|z|^\Delta},
~~~(z<0).
$$
In the next Section, we shall show how the same formula for the one-point function can be obtained from 
the IIB string theory on the $AdS_5\times S^5$ background with a probe D7 brane.   Normally, one would expect that the classical limit
of string theory computes the strong coupling planar limit of the gauge theory, and the computation which we
have performed in the gauge theory is perturbative, valid only at weak coupling.   In this case, however, there are additional
parameters $k_1$ and $k_2$ and in the limit where these parameters are large, there is an effective coupling $\sqrt{\frac{\lambda}{k_1^2+k_2^2}}$
which can be small in both the gauge theory and the classical limit of the string theory.  It is the leading order of the chiral primary one-point
functions in an asymptotic expansion in this parameter that we are comparing.

\section{One-point function from supergravity}

In this section we will obtain the  one-point function  (\ref{onepointfunction})
from a dual supergravity calculation. We begin by reviewing the problem of embedding
a probe D7 brane with internal flux into $AdS_5\times S^5$ and studying the limit where the 
flux is large. 

\subsection{Probe D7 brane}

We shall study the D3-D7 system in the probe limit where the number of D7 branes $N_7$ is much
smaller than $N$, the number of D3 branes.    In principle, with multiple D7 branes, the coordinates are 
matrices and the worldvolume gauge fields have $U(N_7)$ gauge group.  Since the 
non-abelian structure plays no role in this Section, we shall set $N_7=1$ (but of course will need to
restore it in the following sections where we discuss non-Abelian worldvolume gauge fields).   
  We shall be interested in the limit of the string
theory which coincides with the planar limit of the gauge theory and, after the planar limit is 
taken, the large $\lambda$ strong coupling limit.  
In this limit, the string theory is classical, and the problem of including a D7 brane in the $AdS_5\times S^5$ geometry
reduces to that of finding an extremum of the Dirac-Born-Infeld and Wess-Zumino actions,
\begin{align}\label{DBI1}
S=\frac{ T_7}{g_s} \int d^{\:8}\sigma\left[- \sqrt{-\det( g+2\pi\alpha'{\mathcal F})} +\frac{(2\pi\alpha')^2}{2}
C^{(4)}\wedge {\mathcal F}\wedge {\mathcal F}\right], 
\end{align}
where $g_s$ is the closed string coupling constant, which is related to the ${\mathcal N}=4$
Yang-Mills coupling
by $4\pi g_s =g_{YM}^2$, $\sigma^a$ are the coordinates of the D7 brane
worldvolume, $g_{ab}(\sigma)$ is the induced metric, $C^{(4)}$ is the  4-form of the $AdS_5\times S^5$ background,
${\mathcal F}$ is the worldvolume gauge field and
$
T_7=\frac{1}{(2\pi)^7{\alpha'}^4}
$ is the D7 brane tension.  
We shall work with coordinates where  the metric of $AdS_5\times S^5$ is
\begin{align}\label{ads5metric}
ds^2= \sqrt{\lambda}\alpha'&\left[ r^2 (-dt^2+dx^2+dy^2+dz^2) + \frac{dr^2}{r^2}+\right. \nonumber \\
&\left. +d\psi^2+\cos^2\psi(d\theta^2+\sin^2\theta d\phi^2)+ \sin^2\psi (d\tilde\theta^2+\sin^2\tilde\theta d\tilde\phi^2)\right].
\end{align}
Here, $(t,x,y,z,r)$ are coordinates of the Poincare patch of $AdS_5$.  
The boundary of $AdS_5$ is located
at $r\to\infty$ and the Poincare horizon at $r\to 0$.  The coordinates of $S^5$ are identical to those which we used in the previous section. 
The Ramond-Ramond  4-form  is
\begin{align}\label{4form}
C^{(4)}=  \lambda{\alpha'}^2\left[
r^4dt\wedge dx\wedge dy\wedge dz+ \frac{c(\psi)}{2}d\cos\theta\wedge d\phi
\wedge d\cos\tilde\theta\wedge d\tilde\phi\right],
\end{align}
with
$
\partial_\psi c(\psi)=8\sin^2\psi\cos^2\psi
$.
The dynamical variables are
the ten functions of eight worldvolume coordinates which embed the D7 brane in $AdS_5\times S^5$,
as well as the eight worldvolume gauge fields. 
The equations of motion for the embedding can be solved by a worldvolume geometry
 which is $AdS_4\times S^2\times S^2$ and which covers   the whole range of the coordinates
$(t,x,y,r,\theta,\phi,\tilde\theta,\tilde\phi)$.  It sits at a particular value of  $\psi$, which 
must satisfy equation (\ref{equationforpsi}) below, and lies on the curve
\begin{equation}
z = -\frac{\Lambda}{r},\qquad
\Lambda \equiv \frac{f_1f_2}{\sqrt{(f_1^2+4\cos^4\psi)(f_2^2+4\sin^4\psi)-f_1^2f_2^2}}.
\end{equation}
The parameters $f_1$ and $f_2$ are fluxes of the worldvolume gauge fields corresponding to
Dirac monopole on the $S^2$'s, related to
the monopole numbers $k_1$ and $k_2$ by $(k_1,k_2)= \frac{\sqrt{\lambda}}{2\pi}(f_1,f_2)$, so that
\begin{align}\label{backgroundmagneticfield}
{\mathcal F}&= \frac{1}{2}\left(k_1d\cos\theta\wedge d\phi+
k_2d\cos\tilde\theta\wedge d\tilde\phi\right)  \nonumber \\
&= \frac{\sqrt{\lambda}}{4\pi}\left(f_1d\cos\theta\wedge d\phi+
f_2d\cos\tilde\theta\wedge d\tilde\phi\right).
\end{align}
The equations of motion require that the angle $\psi$ solves the equation
\begin{equation}\label{equationforpsi}
(f_1^2+4\cos^4\psi)\sin^2\psi = (f_2^2+4\sin^4\psi)\cos^2\psi,
\end{equation}
The D7 brane worldvolume metric is
\begin{align}\label{D7metric}
ds^2= \sqrt{\lambda}\alpha'\left[ r^2 (-dt^2+dx^2+dy^2)+\frac{dr^2}{r^2}(1+\Lambda^2)+\right. \nonumber \\ \left. +\cos^2\psi(d\theta^2+\sin^2\theta d\phi^2) +\sin^2\psi(d\tilde\theta^2+\sin^2\tilde\theta d\tilde\phi^2)\right],
\end{align}
which is a product of $AdS_4$ with radius of curvature squared $\sqrt{\lambda}\alpha'(1+\Lambda^2)$ 
and two $S^2$'s with radii squared $\sqrt{\lambda}\alpha'$. In the limit where $f_1\gg 1$ and $f_2\gg 1$, 
$$
\Lambda\approx \sqrt{f_1^2+f_2^2}/2 = \frac{\pi}{\sqrt{\lambda}}\sqrt{k_1^2+k_2^2},
$$
and, upon rescaling $r$, the worldvolume metric is
\begin{align}\label{D7metric1}
ds^2=& \frac{\pi^2}{\sqrt{\lambda}}\left(k_1^2+k_2^2\right)\alpha'\left[ r^2 (-dt^2+dx^2+dy^2)+\frac{dr^2}{r^2}\right]+ \nonumber \\ & +\sqrt{\lambda}\alpha'  
\left[\frac{k_1^2}{k_1^2+k_2^2} (d\theta^2+\sin^2\theta d\phi^2) +
\frac{k_2^2}{k_1^2+k_2^2} (d\tilde\theta^2+\sin^2\tilde\theta d\tilde\phi^2)\right].
\end{align}
where we have used the fact that, in this limit,  the angle is given by 
\begin{align}\label{angle}
\tan\psi=\frac{k_2}{k_1},
\end{align}
 which is the same angle as we fixed in the previous section if we identify the number
of units of flux with the dimension of the representation of the SO(3) representations. We see that, in the limit that we are considering, 
the $S^2$'s have radius of curvature squared of order the $\sqrt{\lambda}\alpha'$ of the $AdS_5\times S^5$ background whereas
the $AdS_4$ is much flatter, with radius of curvature squared 
$\frac{\pi^2}{\sqrt{\lambda}}\left(k_1^2+k_2^2\right)\alpha'$. Both
of these must be large in order for the string sigma model to be semi-classical on this background.

 \subsection{The one-point function}

\begin{figure}\begin{center}
 \includegraphics[bb=0 0 570 360, width=5in]{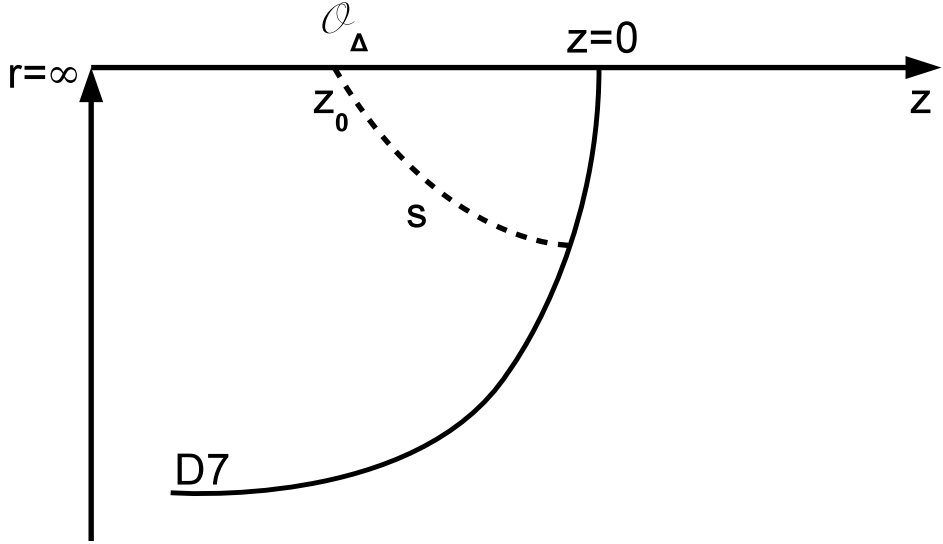}\\
\end{center}
\caption{{\bf Supergravity computation}: The insertion of the
    operator ${\cal O}_\Delta$ on the boundary ($r=\infty$) at $z=z_0$ is dual to a
    supergraviton $s$ which propagates from the boundary insertion
    point into the bulk, where it is integrated over the $D7$ brane
    world volume.}\label{sgfig}
\end{figure}

The calculation involves
computing the fluctuation $\delta S$ of the action of a $D7$ brane due
to fluctuations in the background supergravity fields due to the
insertion of a source on the boundary corresponding to the operator
${\cal O}_\D$ defined in (\ref{chiralprimary}), see Figure \ref{sgfig}. This calculation has
been performed for the case of a $D5$ brane in
\cite{Nagasaki:2012re}; our calculation detailed below proceeds in a
directly analogous manner.

 We will be interested in the large $f_1$ and large $f_2$ limit,
with 
\begin{equation}
\xi \equiv \frac{f_2}{f_1},
\end{equation}
held fixed. In this limit $\psi  = \arctan\xi$. We begin by writing
down the induced metric-plus-worldvolume-field-strength 
\begin{equation}\label{h}
H = 2\pi\alpha'{\mathcal F}+ \p_a X^M \p_b X_M = \sqrt{\lambda}\alpha'
\text{diag}\left(r^2,r^2,r^2,\frac{1+\Lambda^2}{r^2},B_1,B_2\right),
\end{equation}
where
\begin{equation}
B_1 = \begin{pmatrix} 
&\cos^2\psi &-\frac{f_1}{2}\sin\theta\\
&\frac{f_1}{2}\sin\theta &\cos^2\psi\sin^2\theta
\end{pmatrix},~
B_2 = \begin{pmatrix} 
&\sin^2\psi &-\frac{f_2}{2}\sin\tilde{\theta}\\
&\frac{f_2}{2}\sin\tilde{\theta} &\sin^2\psi\sin^2\tilde{\theta}
\end{pmatrix}.
\end{equation}
For simplicity, we will set the factor of $\sqrt{\lambda}\alpha'$ in equation (\ref{h}) to one and 
take it into account by multiplying the D7 brane action by $\lambda^2{\alpha'}^4$.
The inverse of $H$ is given by
\begin{equation}
H^{-1} = \text{diag}\left(r^{-2},r^{-2},r^{-2},\frac{r^2}{1+\Lambda^2},B_1^{-1},B_2^{-1}\right),
\end{equation}
where
\begin{equation}\begin{split}
&B_1^{-1} = \frac{1}{\sin^2\theta(\cos^4\psi+f_1^2/4)}\begin{pmatrix} 
&\cos^2\psi\sin^2\theta &\frac{f_1}{2}\sin\theta\\
&-\frac{f_1}{2}\sin\theta &\cos^2\psi
\end{pmatrix},\\
&B_2^{-1} =  \frac{1}{\sin^2\tilde{\theta}(\sin^4\psi+f_2^2/4)}\begin{pmatrix} 
&\sin^2\psi\sin^2\tilde{\theta} &\frac{f_2}{2}\sin\tilde{\theta}\\
&-\frac{f_2}{2}\sin\tilde{\theta} &\sin^2\psi
\end{pmatrix}.
\end{split}
\end{equation}
The background metric fluctuations $h$ associated to the insertion of the operator at
the boundary give rise to fluctuations of the induced metric
\begin{equation}\label{fluctuations}\begin{split}
{\cal H}_{\alpha\b}= \p_a X^M \p_b X^N h_{MN} =
 \text{diag}\Bigl(h_{tt},h_{xx},h_{yy},h_{rr}+&2\frac{\Lambda}{r^2}h_{zr}
 +\frac{\Lambda^2}{r^4}h_{zz},\\
&h_{\theta\theta},h_{\phi\phi},h_{\tilde{\theta}\tilde{\theta}},
h_{\tilde{\phi}\tilde{\phi}}\Bigr),
\end{split}
\end{equation}
where $h$ and the fluctuation of the background 4-form potential $a$ are given by \cite{Lee:1998bxa} 
\begin{equation}
\begin{split}
&h^{AdS_5}_{\m\n} = -\frac{2\Delta(\Delta-1)}{\Delta+1}g^{AdS_5}_{\m\n} s +
\frac{4}{\Delta+1}\nabla_\m\nabla_\n s,\\
&h^{S^5}_{\a\b} = 2\Delta g^{S^5}_{\a\b}\,s,\\
&a^{AdS_5}_{\m\n\r\s} = 4i\sqrt{g^{AdS_5}} \e_{\m\n\r\s\o}\nabla^\o s,
\end{split}
\end{equation}
where $s$ is a field on $AdS_5\times S^5$, which is decomposed as a
sum of Kaluza-Klein modes on the five-sphere (in the following $\Omega$
denotes coordinates on $S^5$ while $X$ denotes coordinates on $AdS_5$) 
\begin{equation}\label{ssum}
s(X,\Omega) = \sum_{\Delta}\sum_{I} s_{\Delta I}(X)\, Y_{\Delta I}(\Omega),
\end{equation}
where the total spin (i.e. SO(6) Casimir) is represented by $\D$,
while the other quantum numbers defining the state are collected in the index $I$.
Each mode $s_{\Delta I}(X)$ has mass-squared $=\Delta(\Delta-4)$ and
corresponds, via the AdS/CFT dictionary, to a specific chiral primary
operator of dimension $\D$ in the gauge theory, namely to that operator defined by the
spherical harmonic $Y_{\D I}$, where, as in section \ref{sec:g},
the embedding coordinates of the five-sphere are replaced by the six
scalar fields of the gauge theory. Thus, for our purposes, we will be
interested in the specific modes (i.e. single terms in the sum
(\ref{ssum})) corresponding to the operators defined beneath
(\ref{sph}), i.e.\ those with spherical harmonic (\ref{sph}).

We find the following relevant Christoffel symbols ($i$ denotes
coordinates $t,x,y,z$)
\begin{equation}
\G^r_{ii} = -r^3,\qquad
\G^r_{rr} =-\frac{1}{r},\qquad
\G^z_{zr} =\frac{1}{r}.
\end{equation}
The fluctuation of the Euclidean action is then given by 
\begin{align}
\delta S &= \delta S_{DBI} + \delta S_{WZ}  \nonumber \\
&=
\frac{T_7}{g_s}\lambda^2{\alpha'}^4 \int d^8\s \Biggl[\frac{\sqrt{H}}{2} {\rm Tr} \left(H^{-1} {\cal
    H}\right)
+i\frac{(2\pi\alpha')^2}{2}a\wedge{\cal F}\wedge{\cal F}\Biggr] . 
\end{align}
The pull-back of $a$ is
\begin{equation}\begin{split}
&4i\sqrt{g^{AdS_5}} \left(\e_{txyzr} z' r^2 \p_r +
\e_{txyrz}\frac{1}{r^2} \p_z\right)s  \\
&=4i r^3 \left( \Lambda \p_r -\frac{1}{r^2}\p_z\right)s,
\end{split}
\end{equation}
giving
\begin{equation}
\delta S_{WZ} = -\frac{T_7}{g_s} \lambda^2{\alpha'}^4  f_1 f_2 \int d^8\s \,r^3 \sin\theta
\sin\vartheta  \left( \Lambda \p_r -\frac{1}{r^2}\p_z\right)s.
\end{equation}
We find the following expression for the fluctuation of the Dirac-Born-Infeld action
\begin{equation}\begin{split}
&\delta S_{DBI} = \frac{T_7}{2g_s} \lambda^2{\alpha'}^4  \int d^8\s \, r^2 \sqrt{1+\Lambda^2}\sin\theta
\sin\vartheta  \sqrt{(\cos^4\psi+f_1^2/4)(\sin^4\psi+f_2^2/4)} \Biggl\{\\
&\frac{4}{\Delta+1}\left[ \frac{1}{r^2}\left(\p_t^2+\p_x^2+\p_y^2\right) +
    4r \p_r+\frac{r^2}{1+\Lambda^2}\left(\p_r^2
    +2\frac{\Lambda}{r^2}\left(\p_z\p_r-\frac{1}{r}\p_z\right)+
\frac{\Lambda^2}{r^4}\p_z^2\right)\right]\\
&+4\Delta\left(\frac{\cos^4\psi}{\cos^4\psi+f_1^2/4} +
  \frac{\sin^4\psi}{\sin^4\psi+f_2^2/4}\right)
-8\frac{\Delta(\Delta-1)}{\Delta+1} \Biggr\} \,s.
\end{split}
\end{equation}
The field $s$ is replaced by its bulk-to-boundary propagator,
corresponding to a delta-function source $s_0$ on the boundary at 
$((0,0,0,z_0)$
\begin{equation}
s \to  \frac{{Y }_\Delta(\psi) \,c_\Delta}{r^\Delta\left(\rho^2 + (z-z_0)^2 +1/r^2\right)^\Delta},
\end{equation}
where ${Y }_\Delta(\psi)$ is the spherical harmonic (\ref{sph}) (which,
we remind the reader, is not dependent on
$\{\theta,\tilde{\theta},\phi,\tilde{\phi}\}$), while $\rho^2 = t^2+x^2+y^2$, and 
\begin{equation}
c_\Delta \equiv \frac{\Delta+1}{2^{2-\Delta/2}N\sqrt{\Delta}}.
\end{equation}
This normalization is needed to give the bulk two-point function
of the chiral primary unit normalization in the usual supergravity computation \cite{Lee:1998bxa}.
It is also important that we take the operator insertion point at
$z_0<0$. We will also consider the other sign later, and we will find a
different answer. 

The worldvolume integrations over $\theta,\phi,\tilde{\theta},\tilde{\phi}$ are
trivial, as only the volume measure $\sin\theta\sin\tilde{\theta}$ depends
upon them, leading to a factor of $(4\pi)^2$. The integration over
$t,x,y$ are simple radial integrals in the $\rho$ variable. We replace
$dt\,dx\,dy \to 4\pi \r^2d\r$ and use
\begin{equation}\begin{split}
\int_0^\infty d \r \frac{\r^a}{\left(\r^2 + r^{-2}(1+(\Lambda-r z_0)^2)\right)^b} = 
&\left( r^{-2}(1+(\Lambda-r z_0)^2)\right)^{(1+a-2b)/2} \\
&\times\frac{\G\bigl(b-(1+a)/2\bigr)\G\bigl((1+a)/2\bigr)}{2\G(b)}.
\end{split}
\end{equation}
The remaining integral over $r$ is non-trivial, but may be done in
closed form. We then take the $f_1 \to \infty$ limit, keeping
$f_2/f_1$ fixed. It seems that the $\Delta=2$ case is pathological in
that the worldvolume integration over $r$ does not converge. It would
be interesting to understand the interpretation of this pathology
further. On the gauge theory side, there is nothing special about the
$\Delta=2$ case.

\subsection{Results for $z_0<0$}

We first note the following identification of parameters between gauge
and string theory
\begin{equation}
k_i = \frac{\sqrt{\lambda}}{2\pi}\,f_i,
\end{equation}
and then, recalling that $\psi = \arctan(k_2/k_1)$, find the following result
\begin{equation}\label{str}
\begin{split}
\langle {\cal O}_{\Delta}(x)
\rangle=
-\left.\frac{\delta S}{\d s_0}\right|_{s_0=0} =  &
\frac{k_1k_2}{\sqrt{\Delta}}\,
\left(\frac{2\pi^2(k_1^2+k_2^2)}{\lambda}\right)^{\Delta/2}
{Y }_\Delta\bigl(\arctan(k_2/k_1)\bigr)
\,\frac{1}{z_0^\Delta}\\
& + \text{sub-leading in}~ \frac{k_i}{\sqrt{\lambda}},
\end{split}
\end{equation}
where the spherical harmonic $Y_\Delta$ corresponding to the operator ${\cal
  O}_\D$ is given by 
(\ref{sph}) and where $z_0$ should be equated to $z$.
We therefore see that we have obtained a
match with the gauge theory result (\ref{onepointfunction}).

\subsection{Results for $z_0>0$}

Here we find that the leading behaviour of both $\delta S_{DBI}$ and $\delta
S_{WZ}$ goes as
\begin{equation}
\frac{\delta S}{\delta s_0} \sim \lambda^{\Delta/2-2}\,
\frac{k_1k_2}{(k_1^2+k_2^2)^{\Delta/2-2}} \frac{{Y }_\Delta}{z_0^\Delta} 
+ \text{sub-leading in}~ \frac{k_i}{\sqrt{\lambda}}.
\end{equation}
However these leading terms cancel identically between the Dirac-Born-Infeld and Wess-Zumino
parts of the action. Even if they didn't, they are already suppressed
with respect to the $z_0<0$ case for all $\Delta\geq 4$. The cancellation
persists to an additional order in $k_i^2/\lambda$, bringing the
result down by $\lambda^2/k_i^4$, so that for $\Delta=4$, for example,
it goes as $\lambda/k_i^2$. This further cancellation ensures that
$\delta S/\delta s_0$ is suppressed with respect to the $z_0<0$ case for all
$\Delta\geq 2$. This suppression is consistent with the result in the
gauge theory, where we find zero.

\section{$SO(5)$ symmetric solution}

As well as the $SO(3)\times SO(3)$ symmetric configuration of
the D7 brane that we have studied in the previous Sections, there is also
a solution of the D3-D7 brane intersection which has $SO(5)$ symmetry.

\subsection{${\cal N}=4$ Yang-Mills theory defect with $SO(5)$ symmetry}

On the gauge theory side, there is an $SO(5)$ symmetric solution of the Yang-Mills theory
field equations (\ref{fieldequation2})  which represents D3 branes ending on D7 branes in
such a way that an $SO(5)$ subgroup of the $SO(6)$ R-symmetry is preserved.  It has the form
of a fuzzy funnel where, as $|z|$ decreases,  a certain number of D3 branes blow up into a fuzzy 4-sphere,
\begin{align}\label{classicalfoursphere}
\phi_i(x) = \frac{G_i}{\sqrt{8}~z}~~,~~i=1,\ldots,5~,~~~\phi_6=0.
\end{align}
The fuzzy 4-sphere
uses five matrices $G_i$ with $i=1,\ldots,5$ which are the
sum of totally symmetric direct products\footnote{More precisely, operating on a completely
symmetric tensor $v^{\alpha_1\alpha_2\ldots\alpha_n}$, 
$$
G_i v^{\alpha_1\alpha_2\ldots\alpha_n}\equiv \gamma^{\alpha_1}_{~\beta_1}v^{\beta_1\alpha_2\ldots\alpha_n}
+\gamma^{\alpha_2}_{~\beta_2}v^{\alpha_1\beta_2\ldots\alpha_n}+\ldots
+\gamma^{\alpha_n}_{~\beta_n}v^{\alpha_1\alpha_2\ldots\beta_n}.
$$
The dimension of the space of completely symmetric tensors with $n$ indices, with each index running 
from one to four is 
$$
d_G= \frac{(n+3)!}{3! n!}=\frac{1}{6}(n+1)(n+2)(n+3).
$$
}
\begin{align}
G_i=& \left[ \gamma_i\otimes{\mathcal I}_{4\times4}\ldots\otimes {\mathcal I}_{4\times4}
+{\mathcal I}_{4\times4}\otimes\gamma_i\otimes\ldots\otimes{\mathcal I}_{4\times4}
+\ldots +
{\mathcal I}_{4\times4}\otimes\ldots\otimes\gamma_i\right]_{\rm sym} \nonumber \\
   &\oplus{0}_{(N-d_G)\times(N-d_G),}
\end{align}
where $\gamma_i$ are the $4\times 4$ (Hermitian) Euclidean Dirac matrices and each direct product has $n$ factors, the direct sums has $n$ terms
and where sym means restriction to the completely symmetrized tensor
product space.
These matrices are described in
detail in references \cite{Castelino:1997rv}
and \cite{Constable:2001ag}. They have dimension
\begin{align}\label{dg}
d_G=\frac{1}{6}(n+1)(n+2)(n+3),
\end{align}
and
\begin{align}\label{cg}
G_iG_i=c_G~{\cal I}_{d_G\times d_G}\oplus 0_{(N-d_G)\times(N-d_G)}~~,~~c_G(n+4).
\end{align}
Also, the spin matrix
\begin{align}\label{spinmatrix}
G_{ij}=\frac{1}{4}\left[ G_i,G_j\right],
\end{align}
generates $SO(5)$ rotations,
\begin{align}
\left[ G_i,G_{jk}\right] &= \left(\delta_{jk}G_i-\delta_{ik}G_j\right),
\\
\left[ G_{ij},G_{kl}\right] &= \left(
\delta_{jk}G_{il}+\delta_{il}G_{jk}-\delta_{ik}G_{jl}-\delta_{jl}G_{ik}\right).
\end{align}
Now, we can form the classical contribution to the one-point function
of the chiral primary operator by simply plugging the classical solution (\ref{classicalfoursphere}) into 
equation (\ref{chiralprimary})  to obtain,
\begin{align}
\left<{\mathcal O}_\Delta(x)\right> = \frac{d_G}{\sqrt{\Delta}}\left(\frac{\pi^2 c_G}{\lambda}\right)^{\frac{\Delta}{2}}
{\mathcal Y}_{\Delta}(0)\frac{1}{|z|^\Delta},
\end{align}
where ${\mathcal Y}_\Delta(\theta)$ is the $O(5)$ symmetric spherical
harmonic (which we shall derive below),  evaluated at $\theta=0$, the latitude where the
$S^4$ is maximal (since $\phi_6=0$). In the limit where the integer $n$ is large, this reduces
to\footnote{Using the spherical harmonics which will be derived in the next subsecitonb, the 
explicit expressions for the first few values of $\Delta$ are
\begin{align}
\left<{\mathcal O}_2(x)\right> &=-\sqrt{\frac{1}{60}}  \frac{\pi^2 n^5}{\lambda}
\nonumber \\
\left<{\mathcal O}_4(x)\right> &= \sqrt{\frac{9}{11200}} \frac{\pi^2 n^7}{\lambda^2}
\nonumber \\
\left<{\mathcal O}_6(x)\right> &= -\sqrt{\frac{ 1 }{14112 }} \frac{\pi^2 n^9}{\lambda^3}
\nonumber\\
\left<{\mathcal O}_8(x)\right> &= \sqrt{\frac{ 5 }{608256 }} \frac{\pi^2 n^{11}}{\lambda^4}
\nonumber
\end{align}
}
 \begin{align}\label{gres}
\left<{\mathcal O}_\Delta(x)\right> = \frac{n^3}{6\sqrt{\Delta}}\left(\frac{\pi^2 n^2}{\lambda}\right)^{\frac{\Delta}{2}}
{\mathcal Y}_{\Delta}(0)\frac{1}{|z|^\Delta}.
\end{align}

\subsubsection{$O(5)$ symmetric spherical harmonic at level $\Delta$}

Consider the coordinate system where the  metric of the 5-sphere is
$$
ds^2=d\theta^2+\cos^2\theta d\Omega_4^2.
$$
The spherical harmonic that we are interested in does not depend on any of
the coordinates of $S^4$, and therefore satisfies the Laplace equation
$$
\left[\frac{1}{\cos^4\theta}\frac{d}{d\theta}\cos^4\theta\frac{d}{d\theta}+\Delta(\Delta+4)\right]{\mathcal Y}_{\Delta}(\theta)=0,
$$
 which, using the variable
$$
z=\frac{1-\sin\theta}{2},
$$
is the hypergeometric differential equation,
\begin{equation}\label{so(5)hypergeometric}
\left[z(1-z)\frac{d^2}{dz^2} +\left(\frac{5}{2}-5z\right)\frac{d}{dx}+\Delta(\Delta+4)\right]{\mathcal Y}_{\Delta}(z)=0.
\end{equation}
This equation is solved by the hypergeometric function
\begin{equation}
{\mathcal Y}_{\Delta}(z)=C_\Delta ~_2F_1(-\Delta, \Delta+4; \frac{5}{2};z),
\end{equation}
which is a polynomial, 
\begin{align}
_2F_1(-\Delta, \Delta+4; \frac{5}{2};z)=
\sum_{n=0}^\Delta \frac{\Gamma(5/2)}{\Gamma(5/2+n)}
\frac{(\Delta+3+n)!\Delta!}{ (\Delta-n)!(\Delta+3)!n!}
\left(\frac{\sin\theta-1}{2}\right)^n.
\end{align}
Here $C_\Delta$ is a constant which must be fixed so that the spherical harmonic has its
canonical normalization (\ref{normalizationofsphericalharmonic}), 
\begin{align}
C_\Delta = \sqrt{\frac{(\Delta+2)(\Delta+3)}{3\cdot 2^{\Delta+1}}}.
\end{align}
We are interested in these polynomials at $\theta=0$ which can be found
using
Gauss' second summation theorem  
\begin{align}
_2F_1(a,b;\frac{1}{2}(a+b+1);\frac{1}{2})=\frac{\Gamma(\frac{1}{2})\Gamma(\frac{1}{2}(a+b+1))}{\Gamma(\frac{1}{2}(1+a))\Gamma(\frac{1}{2}(1+b))},
\end{align}
to find
\begin{align}
{\mathcal Y}_{\Delta}(\theta=0)=C_\Delta  ~_2F_1(-\Delta,\Delta+4;\frac{5}{2};\frac{1}{2})=
C_\Delta\frac{\Gamma(\frac{1}{2})\Gamma(\frac{5}{2})}{\Gamma(\frac{1}{2}(1-\Delta))\Gamma(\frac{1}{2}(5+\Delta))}.
\end{align}
This vanishes when $\Delta=1,3,5,...$ and, when $\Delta=2\ell$,
\begin{align}
{\mathcal Y}_{2\ell}(\theta=0)=C_{2\ell}  ~
\frac{(\frac{1}{2}-1)(\frac{1}{2}-2)\ldots(\frac{1}{2}-\ell)}
{(\frac{3}{2}+\ell)(\frac{1}{2}+\ell)\ldots(\frac{3}{2}+1)}.
\end{align}
The first few spherical harmonics are
\begin{align}
{\mathcal Y}_0&=1,
\nonumber \\
{\mathcal Y}_1&=\sin\theta,
\nonumber \\
{\mathcal Y}_2&=\sqrt{\frac{1}{30}}\left[5-6\cos^2\theta\right],
\nonumber \\
{\mathcal Y}_3&=\sqrt{\frac{9}{40}}\left[-\sin\theta+\frac{8}{3}\sin^3\theta\right],
\nonumber \\
{\mathcal Y}_4&=\sqrt{\frac{1}{2800}}\left[35-112\cos^2\theta+80\cos^4\theta\right].
\end{align}

\subsection{Supergravity computation}

Now we shall consider the computation of the one-point function using supergravity.  In the scenario where $d_G$ D3 branes end on $N_7$
D7 branes,
they act  as a source of topological charge on the
 world volume of the D7 branes which is reflected in the second Chern class of  the gauge fields on the worldvolume 
of the D7 branes,
\begin{align}\label{instnum}
d_G=\frac{1}{8\pi^2}\int_{S^4} {\rm Tr} F\wedge F,
\end{align}
where the integral is over a 4-cycle in the D7 brane worldvolume which links the D3 brane endpoints.
  
In the large $N$ planar limit and the subsequent large $\lambda$ limit, the problem of embedding the D7 brane in $AdS_5\times S^5$ 
reduces to finding an extremum of the Dirac-Born-Infeld action plus the Wess-Zumino term, 
\begin{align}\label{DBI2}
S=\frac{ T_7}{g_s} \text{STr}\int d^8\sigma\left[ \sqrt{\det(
    g+2\pi\alpha'F)} +i\frac{(2\pi\alpha')^2}{2}
  C^{(4)}\wedge F\wedge F\right], 
\end{align}
where the trace ${\rm STr}$ is computed using the
symmetrized trace prescription  and  we are assuming Euclidean signature of the metric. 
Moreover, the worldvolume gauge fields must take up a configuration such that their second Chern class
is given by equation (\ref{instnum}).  Their appearance in the Wess-Zumino term then influences the geometry of the
D7 brane.  There is a solution of the equations of motion where the D7 brane  is $AdS_4\times S^4$ embedded in
$AdS_5\times S^5$ and the gauge fields are a ``homogeneous instanton'',  which was constructed in reference \cite{Constable:2001ag} and which
has $SO(5)$ symmetry.  This instanton is obtained by simply replacing the Pauli
matrix generators of $SU(2)$ in the usual Belavin-Polyakov-Schwarz-Tyupkin instanton (stereographically projected to $S^4$)
by the $(n+2)$-dimensional irreducible representation of $SU(2)$,
where $n$ is the same integer which appears in equation (\ref{dg}) and then $d_G$ in (\ref{instnum}) is given by (\ref{dg}). This means that the 
number of D7 branes is $N_7+2$.

We shall use a coordinate system where the metric of $AdS_5\times S^5$ is 
\begin{equation}
ds^2= r^2 (-dt^2+dx^2+dy^2+dz^2) +
  \frac{dr^2}{r^2} +d\theta^2+\cos^2\theta d\Omega_4^2,
\end{equation}
where $d\Omega_4^2$ is the metric of the unit 4-sphere. 
The equations for the D7 brane embedding are solved by the space $AdS_4\times S^4$ which covers the space
spanned by the coordinates $(t,x,y,r)$ and wraps  the $S^4\subset S^5$ located at angle $\theta=0$  while the transverse
coordinate is $r$-dependent, 
\begin{equation}\label{definitionofQ}
z = -\frac{\Lambda}{r},\qquad \Lambda \equiv \frac{Q}{\sqrt{1+2Q}}~,
~~~Q=\frac{6\pi^2}{\lambda}\frac{d_G}{n+2}~~.
\end{equation}
The worldvolume metric is
\begin{align}
ds^2=\sqrt{\lambda}\alpha'\left[
r^2(-t^2+x^2+y^2)+ (1+\Lambda^2)\frac{dr^2}{r^2}+
d\Omega_4^2\right].
\label{d7metric2}
\end{align}
We see that, as in the $SO(3)\times SO(3)$ symmetric case, the 
radius of curvature increases as a result of the back-reaction of
the geometry of the D7 brane in the presence of the instanton flux. 
In the limit of large $Q$, it becomes much larger than the radius of
curvature of the $S^4$.  Even though we will shortly consider the case
where $Q$ and $n$ are large, we note that what we have done so far  is valid
for any value of $Q$ with one restriction.  When $Q$ is small, the D7 brane embedding is unstable
to fluctuations of the latitude at which the $S^4$ is located on $S^5$ \cite{Rey:2008zz}. 
It is stable when $Q>\frac{7}{2}$ \cite{Myers:2008me}. We caution the reader that this is a condition
for stability to small fluctuations and it does not rule out the possibility of nonperturbative 
instabilities.

To facilitate our computation, we rewrite the Dirac-Born-Infeld part of the action in the following
way\footnote{See equation (46) of reference \cite{Constable:2001ag}.},
\begin{align}\label{action2}
S_{\text{DBI}} &=\frac{ T_7}{g_s} \text{STr}\int d^8\sigma~\sqrt{\det(
    g+2\pi\alpha'F)} \nonumber \\
&= \frac{T_7}{g_s}\int d^8\sigma
\sqrt{g_{\text{AdS}}}\sqrt{g_S}\left(
(n+2) + \frac{(2\pi\alpha')^2}{4}\text{Tr}\, F_{ab}F^{ab}\right).
\end{align}
This identity is valid when  the D7 brane metric  decomposes into two nonzero $4\times4$ blocks which, on  the 
classical solution,  become $AdS_4$ 
and $S^4$ (as in equation (\ref{d7metric2})).   The metric blocks are therefore denoted $g_{\text{AdS}}$
and $g_{\text{S}}$, respectively.  Moreover, the gauge field has nonzero components only in the $g_{\rm S}$ block
and it is self-dual with respect to the $g_{\rm S}$ geometry,   
$$
F_{ab}=\frac{1}{2}\epsilon_{abcd}F^{cd}.
$$ 
We note that, when the metrics of the blocks are set equal to the $AdS_4$ and $S^4$ metrics 
plus a perturbation by  
the supergraviton field $s(\sigma)$ which corresponds to the chiral
primary operator (the explicit perturbations are $(h^{\text{AdS}},h^{\text{S}^4})$  given in equation (\ref{fluctuations})), they retain 
their block diagonal form and 
\begin{align}
\delta\sqrt{g_{\text{AdS}}} &= \lambda{\alpha'}^2 \sqrt{1+\Lambda^2}r^2\left(1+\frac{1}{2}\text{tr}\,g_{\text{AdS}}^{-1}h^{\text{AdS}}\right), \\
\delta\sqrt{g_S} &=  \lambda{\alpha'}^2\sqrt{g_{S^4}}\left(1+\frac{1}{2}\text{tr}\,g_{\text{S}^4}^{-1}h^{\text{S}^4}\right),
\end{align}  
where $g_{\text{S}^4}$ is the metric of the unit $S^4$.
We  also note that $F_{ab}$ (i.e. with down-indices)
is independent of the metric and therefore is independent of the supergraviton perturbation $s(\sigma)$.
%
%
%
The integrand in equation (\ref{action2}) becomes
\begin{equation}
\begin{split}
\delta {\cal L}_{\text{DBI}} = &\lambda{\alpha'}^2 \sqrt{1+\Lambda^2}r^2\sqrt{g_{\text{S}^4}}\left(
(n+2) +  \frac{(2\pi\alpha')^2}{4}\text{Tr}\,  F_{ab}F^{ab}\right)
\frac{\text{tr}}{2}g_{\text{AdS}}^{-1} h^{\text{AdS}}\\
+&\lambda{\alpha'}^2 \sqrt{1+\Lambda^2}r^2\sqrt{g_{\text{S}^4}}\left(
(n+2) +\frac{(2\pi\alpha')^2}{4}\text{Tr}\,  F_{ab}F^{ab}\right)
\frac{\text{tr}}{2}g_{\text{S}^4}^{-1} h^{\text{S}^4}\\
+&\lambda{\alpha'}^2 \sqrt{1+\Lambda^2}r^2\sqrt{g_{\text{S}^4}}  \frac{(2\pi\alpha')^2}{4}\text{Tr}\,  F_{ab}F^{ab}\left(-\frac{\text{tr}}{2}g_{\text{S}^4}^{-1} h^{\text{S}^4}\right),
\end{split}
\end{equation}
so that the last term cancels part of the second term. 

Now, plugging in the expressions for $(h^{\text{AdS}},h^{\text{S}^4})$  given in equation (\ref{fluctuations})
and noting that $s(\sigma)$ will not depend on the coordinates 
on  $S^4$, integration 
over the $S^4$ yields
\begin{align} 
&\delta S_{DBI} = \frac{T_7}{g_s} (n+2)  \int d^4\sigma \, r^2
    \sqrt{1+\Lambda^2} \, \frac{8\pi^2}{3} \, \frac{1+Q}{2}
\left\{   \frac{4}{\Delta+1}   \left[   \frac{1}{r^2}\left(\partial_t^2+\partial_x^2+\partial_y^2\right) +
\right. \right. 
\nonumber \\
 &\left. \left.  +
    4r \partial_r+\frac{r^2}{1+\Lambda^2}\left(\partial_r^2
    +2\frac{\Lambda}{r^2}\left(\partial_z\partial_r-\frac{1}{r}\partial_z\right)+
\frac{\Lambda^2}{r^4}\partial_z^2\right)  \right] 
-8\frac{\Delta(\Delta-1)}{\Delta+1}  \right\}  \,s(\sigma)   \nonumber\\
&+\frac{T_7}{g_s} (n+2)  \int d^4\sigma \, r^2
    \sqrt{1+\Lambda^2}\,\frac{8\pi^2}{3}  \,4\Delta s(\sigma).
\nonumber \end{align}
where the unit 4-sphere volume is $\frac{8\pi^2}{3} $ and 
$Q$ is defined in equation (\ref{definitionofQ}).
The last term stemming from the remaining variation in
$g_{S}$ is sub-leading, since it does not have a factor of $Q$
associated to it. 

The Wess-Zumino term in the action has a structure   similar to 
that in the two-flux case
that we studied in the previous section, as $s(\sigma)$ appears 
only in the Ramond-Ramond potential, so that
\begin{equation}
\delta S_{WZ} = -\frac{T_7}{g_s} \frac{32\pi^2}{3}(n+2) Q  \int
d^4\sigma \,r^3 \left( \Lambda \partial_r -\frac{1}{r^2}\partial_z\right)s(\sigma),
\end{equation}
where we have used (\ref{instnum}). The remainder of the problem simply
follows the procedure of the 
two-flux case that we used in the previous section.  We find that, when  $z_0<0$, and in the limit
where $n$ and $Q $ are large, 
\begin{equation}
\langle{\mathcal O}_\Delta(x)\rangle=\frac{1}{\sqrt{\Delta}} \frac{n^3}{6z_0^\Delta} 
\left(\frac{\pi^2n^2}{\lambda}\right)^{\Delta/2} \, {\cal Y}_{\Delta}(0),
\end{equation}
which is identical to the Yang-Mills theory result (\ref{gres}).  

\subsubsection{$z_0>0$}

It is also easy to show that, on the other side of the D7 brane, where $z_0>0$, 
\begin{equation}
\langle{\mathcal O}_\Delta(x)\rangle \sim Q^{1-\Delta/2},
\end{equation}
is much smaller than that on the $z_0<0$ side.  This is consistent
with the fact that, in the gauge theory,
the leading order result vanishes.

\section{Discussion}

We have computed the one-point functions of chiral primary operators
in two separate cases of the D3-D7 brane system.  Both computations
can be done in the gauge theory and in the string theory dual and, in
a particular large flux limit, the computations can be compared and
they indeed do match. It would be very interesting to check this
agreement beyond the leading order that we have computed.  In
principle, the string theory computation can be done for any value of
the monopole or instanton numbers.  Accuracy of the computation
depends only on $\sqrt{\lambda}$ being large.  It would thus be easy
to find the sub-leading orders of the computation.  On the gauge
theory side, on the other hand, the sub-leading orders would require
computing perturbation theory in small $\lambda$.  In both cases,
however, one could generate an asymptotic series in
$\sqrt{\lambda}/k$.  It could happen that, like the BMN limit
\cite{Berenstein:2002jq} of the string spectrum, such a comparison
could fail at a higher order simply due to a non-commutativity of
limits problem \cite{Callan:2003xr}.  It would be very interesting to investigate this.

A computation of this kind could be helpful in constructing the quantum field theory on the defect.   The bi-fundamental
fields living on the defect  should be chiral fermions.  Local self-interactions of fermions are irrelevant dimension four and greater
operators.  A candidate Lagrangian for the fermions is
$$
S_{\rm defect}=\int d^3x\left\{ \bar\psi(x)i\gamma^\mu (\partial_\mu -i A_\mu)\psi(x) +\ldots\right\},
$$
where $A$ is the bulk gauge field and its coupling to the defect fermions is fixed by gauge invariance.  With a single
D3 brane, the fermions are doublets of the SO(2,1) Lorentz symmetry and 
singlets of $SO(3)\times SO(3)$, so the above are apparently the only marginal operators which contain the defect fermions.    
We are also allowed to add operators made only from the bulk fields and comparison
with the string theory could be a be a useful guide as to what these operators are.  There is a possibility that they could
also be checked by requiring that the resulting theory has conformal symmetry. 

Finally, it would be interesting to examine similar issues for other
defect theories, a good example being the theory based on the Janus
solution \cite{Bak:2003jk,D'Hoker:2006uu,D'Hoker:2006uv}, for which
fully back-reacted backgrounds are also known
\cite{Bachas:2011xa,Assel:2011xz}.

\section*{Acknowledgments}

C.F.K.\ and D.Y.\ were supported in part by FNU through grant number
272-08-0329.  G.W.S. is supported by NSERC of Canada and by the Villum
foundation through their Velux Visiting Professor program. In
addition, G.W.S.  acknowledges the kind hospitality of the Niels Bohr
Institute and the University of Copenhagen. The authors thank the
organizers of the workshop ``The Holographic Way'' held at Nordita,
Stockholm, where this work was completed.

\end{document}